**Selective Actuation Enabled Multifunctional Magneto-mechanical Metamaterial for Programming Elastic Wave Propagation**

*Jay Sim, Shuai Wu, Sarah Hwang, Lu Lu, Ruike Renee Zhao**

J. Sim, S. Wu, S. Hwang, L. Lu, Prof. R. R. Zhao

Department of Mechanical Engineering, Stanford University, Stanford, CA 94305, USA

E-mail: rrzhao@stanford.edu

Active metamaterials are a type of metamaterial with tunable properties enabled by structural reconfigurations. Existing active metamaterials often achieve only a limited number of structural reconfigurations upon the application of an external load across the entire structure. Here, we propose a selective actuation strategy for inhomogeneous deformations of magneto-mechanical metamaterials, which allows for the integration of multiple elastic wave tuning functionalities into a single metamaterial design. Central to this actuation strategy is that a magnetic field is applied to specific unit cells instead of the entire metamaterial, and the unit cell can transform between two geometrically distinct shapes, which exhibit very different mechanical responses to elastic wave excitations. Our numerical simulations and experiments demonstrate that the tunable response of the unit cell, coupled with inhomogeneous deformation achieved through selective actuation, unlocks multifunctional capabilities of magneto-mechanical metamaterials such as tunable elastic wave transmittance, elastic waveguide, and vibration isolation. The proposed selective actuation strategy offers a simple but effective way to control the tunable properties and thus enhances the programmability of magneto-mechanical metamaterials, which also expands the application space of magneto-mechanical metamaterials in elastic wave manipulation.

## 1. Introduction

Active metamaterials[1,2] have recently garnered much research interest because of their tunable properties in response to external stimuli such as mechanical loads,[3] thermal loads,[4] electrical current,[5] and magnetic fields.[6] Compared to traditional metamaterials which have a prescribed unit cell design and can only achieve specific properties once they are fabricated, the unit cells of active metamaterials can reconfigure between different states under actuation and thus achieve property tuning. Among the different actuation strategies, mechanical, thermal, and electrical loads are usually tethered, as they require metamaterials to be in direct contact with the actuation source. For example, mechanical load-driven reconfigurable metamaterials often require physical manipulation from loading equipment[3,7–9] or manual operation by hand[1,10–13]; pneumatic metamaterials need tubing for pressure control[14,15]; many thermal-responsive metamaterials rely on conductive wires to induce joule heating[16–18] or must be in direct contact with a heating apparatus[19–21]; and electrical current-driven active metamaterials also require a conductive medium for current flow.[5,22,23] Although these structural reconfiguration methods can be implemented into an active metamaterial system, tethered actuation often interferes with the deformation, which constrains the structural reconfiguration of the metamaterial. Although some untethered thermal actuation strategies such as convection heating[24,25] and photo-heating[26–28] have been explored, which allow for actuation across distances, they often require significant time in both heating and cooling cycles, which is not ideal for application scenarios that demand rapid structural reconfigurations. Magnetic fields, on the other hand, have demonstrated untethered and ultrafast (sub-second) shape change capabilities,[29–31] making it a robust actuation method for reversible structural reconfiguration of active metamaterials. Recent studies have utilized magnetic actuation to tune the mechanical properties of metamaterials, allowing for tunable stiffness[32] and Poisson's ratio.[33,34] Furthermore, the reconfigurable shapes achieved by magnetic actuation have been explored for elastic wave manipulation, where changes in the metamaterial structure affect the elastic wave bandgap.[35,36] Similarly, magnetically actuated metamaterials have been studied for tunable electromagnetic properties, where different unit cell designs achieve various electromagnetic filter behaviors.[37]

Although the structural reconfiguration of active metamaterials has demonstrated promising capability in property tuning and wave manipulation, most actuation methods only induce homogeneous shape transformations with a limited number of unique configurations.[35–37] To enable versatile structural programmability for enhanced property tunability and multifunctional wave manipulation, inhomogeneous shape transformation through selective actuation of metamaterial unit cells can offer significant advantages over homogeneous shape change.[38–40] For example, the metamaterial can be tailored at different locations within the structure and opens new possibilities for functionalities and applications. In recent years, several works have employed selective actuation to program the mechanical behavior[41–43], phase transitions[44,45], and elastic wave behavior.[46,47] Among different selective actuation strategies, magnetic actuation shows superior capability to easily program the local shape reconfiguration for mechanical property tuning,[42] and certain wave tuning.[44,46] However, in terms of wave tuning, existing works mostly focus on single wave manipulation functions, such as programming the speed and direction of 1D transition waves[44] or utilizing selective actuation as elastic waveguides.[46]

In this work, we report a selective actuation strategy that can induce inhomogeneous deformation of magneto-mechanical metamaterials, which allows for significant structural configuration tunability and thus enables the integration of multiple wave tuning functionalities into a single metamaterial design, for switching between functions such as transmittance tuning, waveguide, and vibration isolation. The selective actuation is achieved by individually actuating the unit cells of the magneto-mechanical metamaterial via localized magnetic field provided by permanent magnets, allowing the metamaterial to be programmed into patterns for various inhomogeneous and unique configurations that are unattainable by conventional actuation methods. Each of the selectively actuated configurations exhibits distinctive wave tuning, such as elastic wave transmittance behaviors, enabling precise fine-tuning of transmittance across multiple frequency ranges. We further demonstrate that by selectively actuating the metamaterial into different inhomogeneous structural configurations, the metamaterial can achieve elastic waveguide, programmable vibration isolation, and direction-dependent elastic wave propagation control. Moreover, dynamic selective actuation in a single-layer and static selective actuation in a bilayer magneto-mechanical metamaterial are presented to show the versatility of the proposed actuation strategy. Overall, the selective actuation strategy significantly enhances the programmability of magneto-mechanical metamaterials in elastic wave manipulation and can be extended to other multifunctional applications.

## 2. Results

### 2.1. Selective actuation via magnetic programming

We start by introducing the unit cell design of the magneto-mechanical material studied in this work. It is designed by following three basic design principles: repeatable patterning in horizontal and vertical directions, ease of fabrication and magnetization, and the ability to undergo significant shape change after actuation. As illustrated in **Figure 1A**, the unit cell consists of eight magnetized rectangular blocks connected by thin joints and is made of hard-magnetic soft materials.[48,49] The unit cell is fabricated and magnetized as a single part, with the magnetization directions of the blocks indicated by the white arrows. In the absence of an applied magnetic field, the unit cell is in the deployed mode, and its left and right sides slightly bend inwards due to magnetic repulsion from neighboring unit cells in the horizontal direction, while its upper and lower sides are slightly bend outwards due to magnetic attraction from adjacent unit cells in the vertical direction. Upon the application of a downward magnetic field, the magnetization direction of each block tries to align itself with the external magnetic field, and the unit cell deforms into a cross-shaped configuration, defined as the folded mode. The folded mode switches back to the deployed mode once the applied magnetic field is removed. Such a unit cell design brings an anisotropic elastic wave propagation behavior. To illustrate this, **Figure 1B** shows the wave propagation in the deployed unit cell along the horizontal and vertical directions. Although both input waves (blue) exhibit the same frequency, the amplitude ($A_1$ and $A_2$) of the output waves (pink) are not equal due to the different distances and orientations of the magnetized rectangular blocks along the two directions. Note that this anisotropic behavior in elastic wave propagation is present in both the deployed and folded modes.

To trigger the transformation between the two modes, we design an actuation magnet cell by installing an axially magnetized neodymium-iron-boron (NdFeB) cylindrical magnet inside a 3D printed casing, as illustrated in **Figure 1C**. The casing is placed directly underneath the substrate on which the unit cell and overall metamaterial rests. Additional information on the actuation magnet cell is provided in Figure S1 (Supporting Information). By placing actuation magnet cells into the corresponding spots where the metamaterial unit cell needs to be actuated, the metamaterial array can be selectively actuated to form a target pattern. The effect of distance between the magnets and the metamaterial, which determines the magnitude of the magnetic field applied for unit cell deformation, is shown in Figure S2 (Supporting Information). **Figure 1D** shows an example to illustrate this concept, in which the target "S" pattern is digitized into a grid of two types of pixels. The first pixel, colored in navy, corresponds to no magnetic actuation, and is conceptualized as a "0" bit. The second pixel, colored in gray, represents a grid position with magnetic actuation and is considered as the "1" bit. The physical actuation magnet cells are plugged into a pegboard to match the target pattern, with the metamaterial placed on top to achieve selective actuation. By independently switching each unit cell between "0" bit and "1" bit, the target pattern can be programmed. The experimental demonstration of various selectively actuated target patterns is presented in Video S1 (Supporting Information). By harnessing the selective actuation strategy, magneto-mechanical metamaterials enable many functionalities that cannot be achieved through global actuation such as tunable elastic transmittance and guided elastic wave control. **Figure 1E** shows a schematic of how actuating different unit cells can lead to functionality change of metamaterial. When selectively actuating the centermost nine unit cells of an $8 \times 8$ magneto-mechanical metamaterial array to the folded mode, the metamaterial changes to a different geometry and its elastic wave transmittance also changes accordingly. When switching the three middlemost columns of unit cells to the folded mode (**Figure 1E** right), the metamaterial behaves as a wave director with the folded unit cells guiding the horizontal elastic wave to propagate along the vertical direction. In general, selective actuation enables inhomogeneous deformation, which in turn, unlocks different shape configurations and their corresponding functionalities.

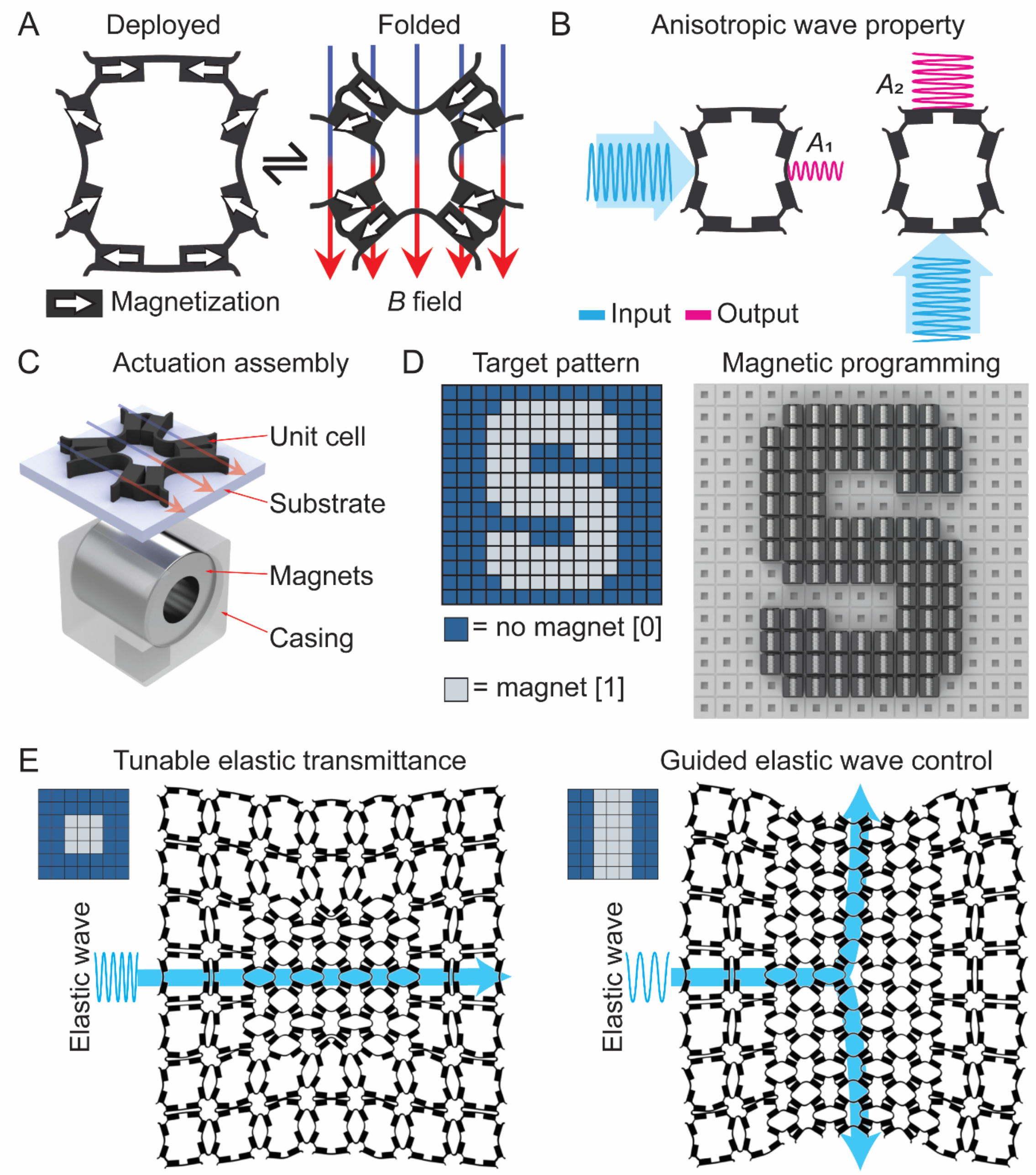

**Figure 1.** Selective actuation of magneto-mechanical metamaterials for tunable elastic wave properties. (A) The magnetic actuation of a unit cell between the deployed and folded modes (B) A wave that travels horizontally through the unit cell results in a transmitted amplitude of $A_1$, which is different than the transmitted amplitude of a wave of the same frequency travelling vertically, $A_2$. This demonstrates the anisotropic wave property of the unit cell. (C) Application of local magnetic field for selective actuation. An actuation magnet cell is placed directly underneath a unit cell. (D) Magnetic programming for selective actuation. Actuation magnet cells are first manually programmed onto a pegboard to match the target pattern and are then positioned underneath the magneto-mechanical metamaterial to selectively actuate the corresponding unit cells. (E) Different selectively actuated patterns for wave transmittance tuning and guiding elastic wave propagation.

### 2.2 Selective actuation for inhomogeneous deformations and transmittance tuning

**Figure 2** demonstrates how selective actuation can be used to tune the elastic wave transmittance of magneto-mechanical metamaterials. Here, we consider six different selectively actuated patterns of an 11 × 11 metamaterial array, as shown in **Figures 2A-F**, which are referred to as the 3 middle rows pattern, the outer rows pattern, the diagonal pattern, the center 7 × 5 pattern, the corners pattern, and the triangle pattern, respectively (an example of homogenous global actuation is demonstrated in Figure S3 in Supporting Information). The distance between the bottom of the metamaterial and longitudinal axis of the actuation magnet cell is 9 mm (Further information can be found in Figure S2, Supporting Information). For the 3 middle rows pattern, the unit cells in the three middle rows of the array are actuated (**Figure 2A**). In the outer rows pattern, the unit cells in the two top rows and two bottom rows are actuated (**Figure 2B**). Unit cells along a diagonal line from the lower left corner to the upper right corner are actuated in the diagonal pattern (**Figure 2C**). For the center 7 × 5, corners, and triangle patterns, the unit cells in a 7 × 5 grid in the center of the array (**Figure 2D**), in a 3 × 3 grid in each corner of the array (**Figure 2E**), and in a triangular grid in the center of the array (**Figure 2F**) are actuated, respectively. For each of the selectively actuated patterns, both the structural finite element analysis (FEA) prediction and experimental validation are depicted. In the structural FEA predictions, the outer boundary of the metamaterial array is free. Figure S4 (Supporting Information) shows a comparison of selective actuation with free and fixed boundary conditions. For practical applications where a globally fixed boundary condition is required, the bilayer strategy detailed in **Figure 6** can be adopted. The magnetic field arrow next to the selective actuation patterns in **Figures 2A – F** illustrates the direction of the applied local magnetic fields. Details on the magnetic actuation simulation are provided in Figure S5 (Supporting Information). It is seen from **Figure 2** that for all patterns, there is excellent agreement between the FEA and experiments. Importantly, after at least 200 cycles of actuation, the unit cells remain intact, demonstrating their reliability through repeated selective actuation. Further information can be found in Figure S6 (Supporting Information).

The final deformed states of the different selectively actuated patterns of the magneto-mechanical metamaterial exhibit distinct wave transmittance. **Figure 2G** depicts the transmittance curves of the outer rows, triangle, and diagonal patterns of the metamaterial predicted by FEA. From 1250 to 2500 Hz, there are three highlighted frequency ranges of interest. In the first highlighted range (with frequency near 1250 Hz), the transmittance of the outer rows and triangle patterns are close to 0 dB, meaning no loss, while that of the diagonal pattern is around -150 dB. This suggests that elastic waves in this frequency range pass through the outer rows and triangle patterns but are blocked by the diagonal pattern. In the second and third ranges (with frequencies near 1500 Hz and 2250 Hz, respectively), the outer rows pattern shows a much higher transmittance than the other two patterns, indicating that the outer rows pattern allows the elastic wave propagation while the triangle and diagonal patterns block the elastic wave propagation. These results also demonstrate that the elastic wave transmittance of the magneto-mechanical metamaterial can be tuned within a wide frequency range through the selective actuation strategy, which has great potential in advanced elastic wave control applications.

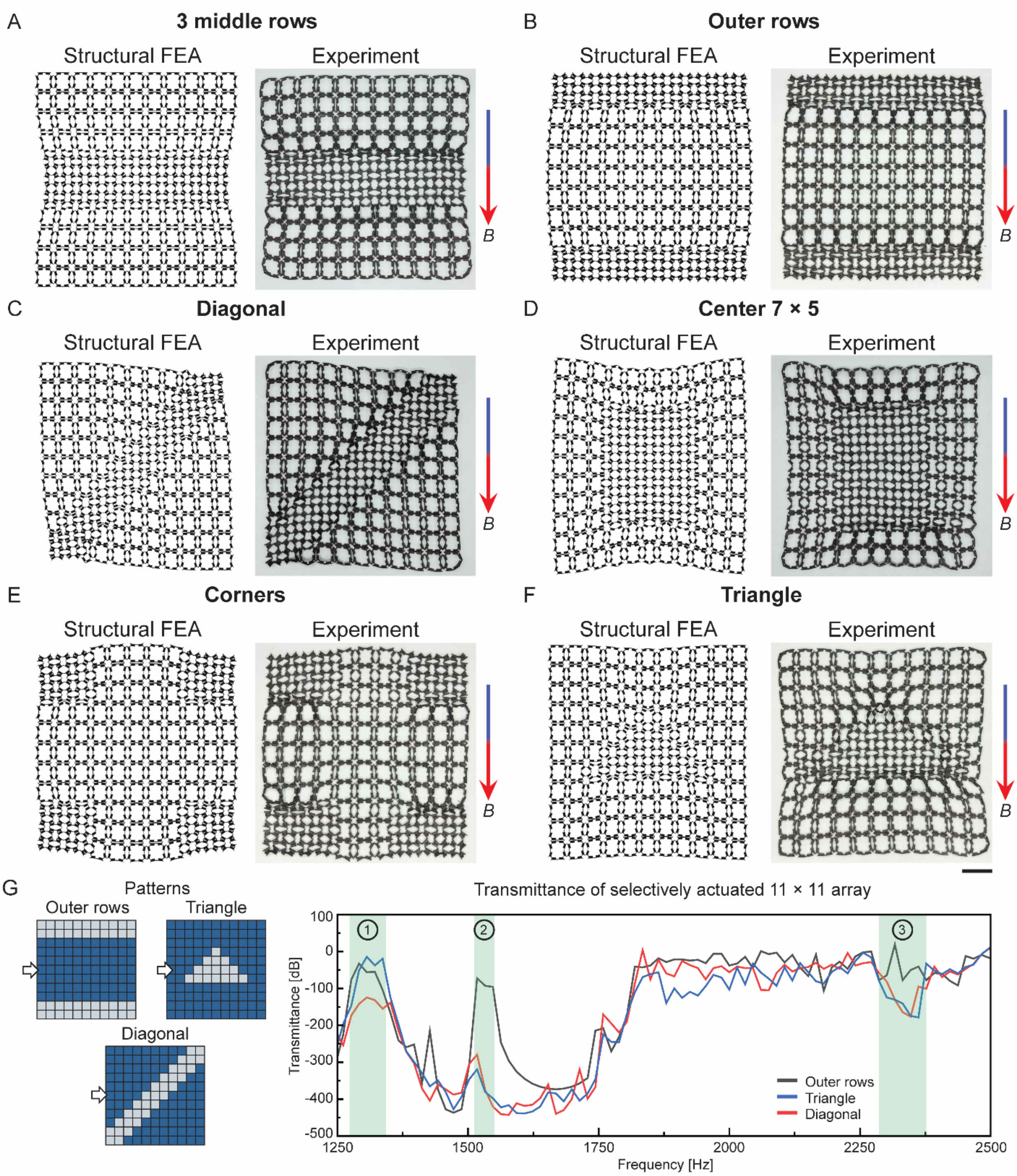

**Figure 2.** Selective actuation of magneto-mechanical metamaterial for elastic wave transmittance tuning. (A-F) Structural FEA prediction and experimental validation of various selectively actuated patterns (A) 3 middle rows pattern; (B) outer rows pattern; (C) diagonal pattern; (D) center 7 × 5 pattern; (E) corners pattern; (F) triangle pattern. Scale bar: 20 mm. (G) Transmittance curves of the outer rows, triangle, and diagonal patterns predicted by FEA with three highlighted frequency ranges of interest.

### 2.3. Elastic wave propagation control for waveguides and vibration isolation

Selective actuation also unlocks additional capabilities for a single magneto-mechanical metamaterial, such as elastic waveguides and vibration isolation. **Figure 3A** shows the transmittance curves of a fully deployed and a fully folded magneto-mechanical metamaterial within a frequency range of 0 and 350 Hz. Two significantly different transmittance-frequency responses are observed for the deployed and folded modes. At 100 Hz, the transmittance of the deployed mode is approximately -50 dB while the folded mode exhibits a much higher transmittance at -1 dB. In comparison, the deployed mode's transmittance is significantly higher than that of the folded mode at 273 Hz. Consequently, by utilizing the varying elastic wave transmittances between the deployed and folded modes, effective waveguides can be achieved through selective actuation. To demonstrate this, **Figure 3B** compares the elastic wave propagation in a 3 middle rows pattern and a fully deployed pattern of an 11 × 11 metamaterial array simulated by FEA (see Supporting Information for details on the elastic wave propagation simulation). In the two cases, an excitation with frequency of 100 Hz is applied to the unit cell in the middle row of the array with the excitation positions marked by red dots. The wave propagation is visualized with a displacement magnitude contour normalized by the initial excitement amplitude, $U / U_0$ (Video S2, Supporting Information). In the selectively actuated 3 middle row pattern, although there are small displacements in the deployed regions, the most significant displacements occur within the folded regions. This indicates that the elastic wave mainly propagates in the folded unit cells along the horizontal direction. In comparison, in the fully deployed pattern, the elastic wave fails to propagate because its unit cells feature very low transmittance at 100 Hz.

In addition, the extent of the elastic wave propagation can be controlled through selective actuation (Video S3, Supporting Information). **Figure 3C** presents the elastic wave propagation in a selectively actuated pattern of a 3 × 12 metamaterial array where the left half of the metamaterial is deployed, and the right half is folded. An excitation with a frequency of 100 Hz is applied to the leftmost unit cells and the excitation positions are denoted by the red dots. Due to the low transmittance of elastic waves in the deployed mode at 100 Hz, the initial excitation is blocked by the left half unit cells. In comparison, in the fully folded pattern, as shown in **Figure 3D**, the 100 Hz elastic wave successfully propagates throughout the entire structure. However, if the excitation frequency is increased to 273 Hz, as depicted in **Figure 3E**, the selectively actuated 3 × 12 metamaterial array acts as a vibration isolator. In this case, the deployed unit cells in the left half of the pattern, which have high transmittance at 273 Hz, allow wave propagation but the folded unit cells in the right half of the pattern, which have low transmittance at 273 Hz, obstruct and isolate the wave propagation. This ensures the wave only passes through half the metamaterial, demonstrating vibration isolation behavior. It is important to note that mechanical interaction between the folded and deployed unit cells causes some shape change in the deployed unit cells that are adjacent to the fully folded actuated unit cells. However, the wave still transmits fully in these imperfect deployed unit cells, demonstrating that the metamaterial's wave propagation property is insensitive to the small shape perturbations. For a fully folded metamaterial excited at 273 Hz, as seen in **Figure 3F**, the wave propagation is blocked because all unit cells are folded and have low transmittance at this frequency. In addition, isolating the propagation to only part of the metamaterial is not possible as all the unit cells in the metamaterial are of the folded mode and therefore have the same frequency-dependent wave properties.

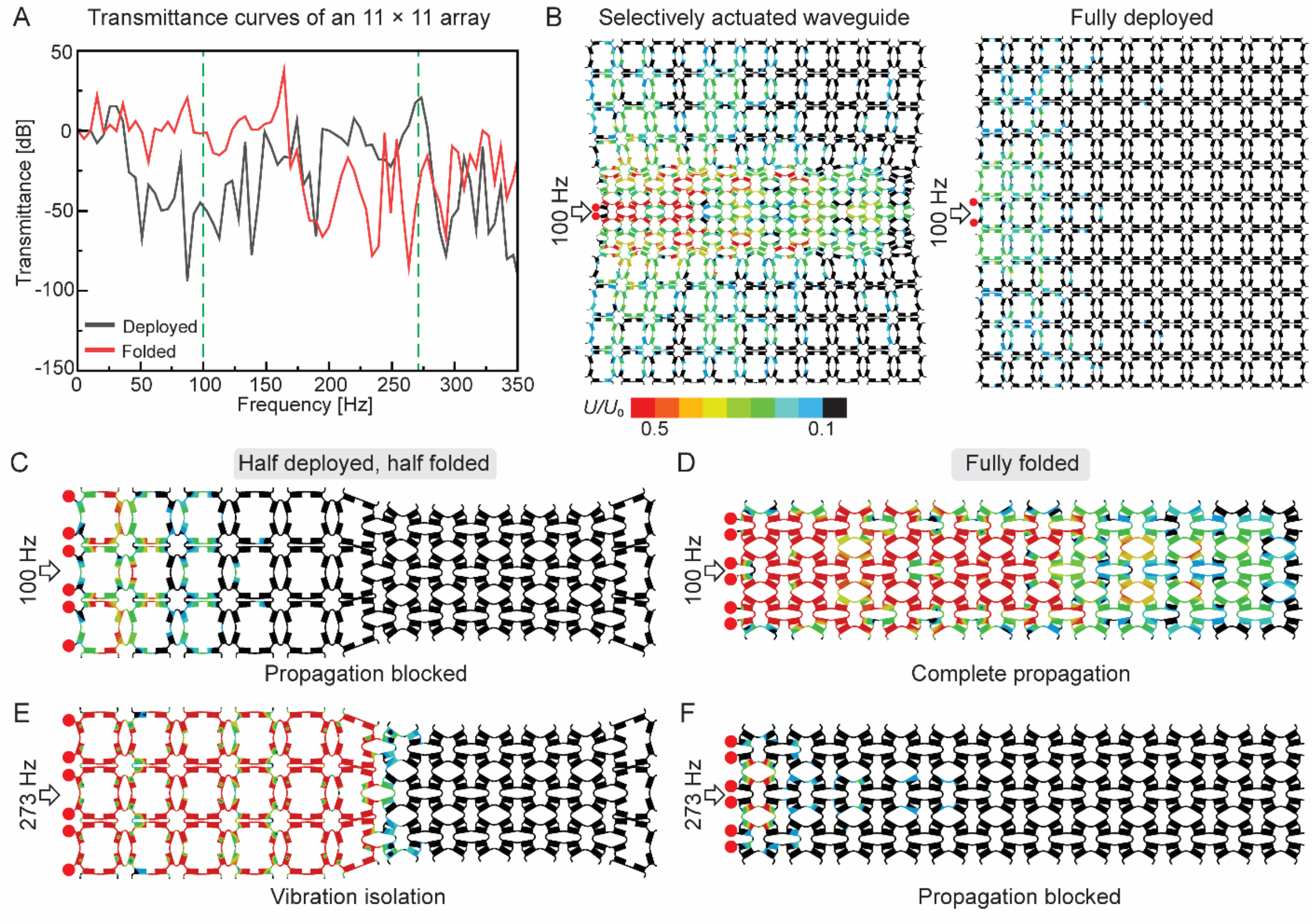

**Figure 3.** Selective actuation of magneto-mechanical metamaterial for waveguides and vibration isolation. (A) Transmittance curves of a fully deployed and fully folded 11 × 11 array from 0 to 350 Hz. Two frequencies of interest, 100 Hz and 273 Hz, are marked by dashed green lines. (B) Wave propagation in the selectively actuated 3 middle rows pattern and the fully deployed pattern for an 11 × 11 array. The red dots denote the location of the initial excitement. (C) Elastic wave blocking on a selectively actuated 3 × 12 metamaterial array with left half unit cells deployed and right half unit cells folded under a 100 Hz excitation and (D) its fully folded pattern comparison. (E) Vibration isolation in a selectively actuated 3 × 12 array with left half unit cells deployed and right half unit cells folded under a 273 Hz excitation and (F) its fully folded pattern comparison.

### 2.4. Direction-dependent elastic wave propagation control

As noted, the unit cell of the magneto-mechanical metamaterial shows an anisotropic wave propagation property in the horizontal and vertical directions. In this subsection, we further demonstrate that the underlying property integrated with selective actuation can be utilized to achieve direction-dependent elastic wave control. **Figure 4A** plots the transmittance curves in the horizontal and vertical directions of the fully deployed pattern (left) and the fully folded pattern (right) of an 11 × 11 metamaterial array simulated by FEA. It is seen that within the frequency range between 0 to 500 Hz, the deployed pattern shows different transmittance curves in the horizontal and vertical directions, such as at 100 Hz and 370 Hz, as marked by the dashed green lines. Likewise, the folded pattern has different transmittance curves in the horizontal and vertical directions, but at 370 Hz, the transmittance is similar in both directions. The wave propagation in the fully deployed pattern when applying a 100 Hz excitation in the horizontal direction is depicted in **Figure 4B**, with the red dots denoting the excitation positions. At this frequency and direction, which are the same as the propagation example depicted in **Figure 3B,** the elastic wave is blocked, and no significant displacement is observed throughout the metamaterial. Conversely, when the same fully deployed pattern is excited vertically, as shown in **Figure 4C**, with the same number of excited unit cells and total number of excitations, there is full elastic wave propagation throughout the metamaterial. Thus, the metamaterial demonstrates obvious direction-dependent wave propagation behaviors at certain frequencies, and this property can be integrated with selective actuation to achieve more diverse wave propagation control. **Figure 4D** presents the wave propagation in the vertical direction of an 11 × 11 metamaterial array which is selectively actuated such that the upper seven rows are folded, and the lower four rows are deployed. As illustrated by the red dots, three unit cells along the lower edge are vertically excited at 370 Hz. At this frequency, the transmittance is high in the vertical direction and low in the horizontal direction for the deployed pattern. Therefore, in the deployed region, the elastic wave mainly propagates along the vertical direction. However, once the wave reaches the folded region, it begins to propagate in both the horizontal and vertical directions because the folded pattern excited at 370 Hz exhibits high transmittance in both directions. As a result, the elastic wave, whose propagation is marked by a dashed red line, propagates in a diffusion manner from the deployed to folded regions. In comparison, a different selectively actuated pattern unlocks the elastic wave directing capability of the metamaterial, as shown in **Figure 4E**. Here, the actuation is inversed, with the upper seven rows being deployed and the lower four rows folded. In this case, an excitation with frequency of 370 Hz is applied to five unit cells along the lower edge vertically. The elastic wave propagates both vertically and horizontally in the folded region as seen by the displacement contour and dashed red boundary. However, once the propagation reaches the interface between folded and deployed regions, the direction-dependent property of the unit cell causes the wave to propagate only vertically in the deployed region with a reduced magnitude. Therefore, the elastic wave is directed vertically from the folded to deployed regions. This direction-dependency in elastic wave propagation is also shown in Video S4 (Supporting Information). These results demonstrate elastic waves can be guided along specific directions in one region of the metamaterial and upon reaching another region, the waves are either spread out and diffused or directed along a secondary direction. Therefore, a single metamaterial can be selectively actuated into different configurations that, when combined with direction-dependent propagation property, enable additional wave manipulation capabilities.

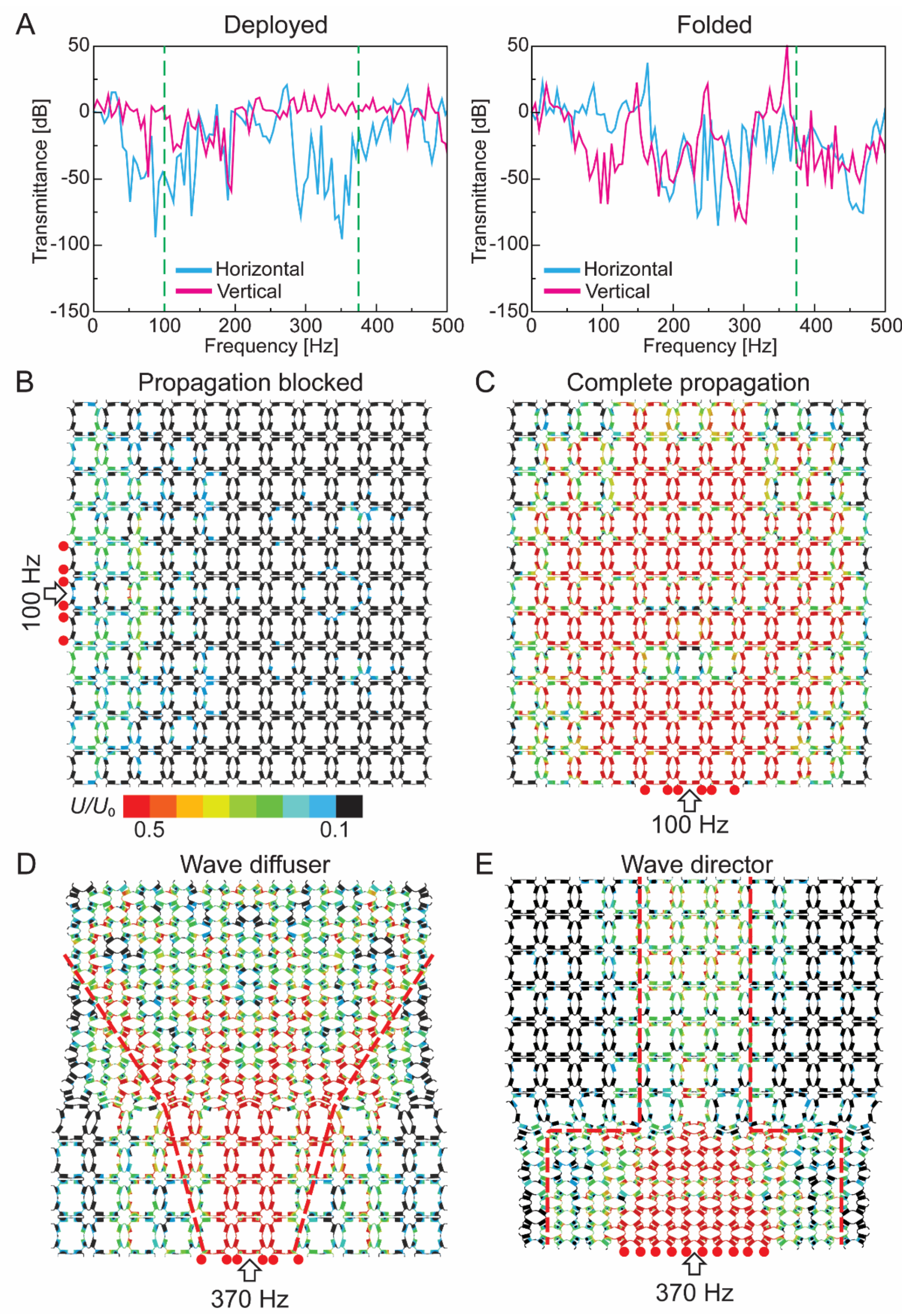

**Figure 4.** Selective actuation of magneto-mechanical metamaterial for direction-dependent elastic wave propagation control. (A) Elastic wave transmittance curves of both the fully deployed and fully folded pattern of an 11 × 11 metamaterial array in the horizontal and vertical directions. Frequencies of interest, 100 Hz and 370 Hz for the deployed pattern and 370 Hz for the folded

pattern, are marked by dashed green lines. (B, C) Elastic wave propagation in a fully deployed 11 × 11 metamaterial array under an excitation frequency of 100 Hz (B) in the horizontal direction and (C) in the vertical direction. (D) Elastic wave propagation, surrounded by a dashed red line, in a selectively actuated 11 × 11 metamaterial array with upper seven rows folded and lower four rows deployed under a 370 Hz excitation in the vertical direction. (E) Elastic wave propagation, surrounded by a dashed red line, in a selectively actuated 11 × 11 metamaterial array with the upper seven rows deployed and lower four rows folded under a 370 Hz excitation in the vertical direction.

### 2.5. Dynamic magnetic selective actuation

In the previous subsections, inhomogeneous shape change is achieved through selective actuation, where the actuation magnet cells responsible for shape change remain in place and thus the actuated patterns are static. Here, we demonstrate that the inhomogeneous shape change can also be dynamic as we move the actuation magnet cells along a defined path. **Figure 5A** shows an illustration of the magnetic programming for translating the actuation magnet cells. Initially, nine actuation magnet cells arranged in a 3 × 3 grid are positioned in the middle of the lower edge of the metamaterial. They then translate upward until reaching the upper edge of the metamaterial, following the direction of the green arrow. The experimental actuation is shown in **Figure 5B** as well as Video S5 (Supporting Information). Along the translation path, four positions of interest are marked 1 to 4. At position 1, the actuation magnet cells are positioned at the bottom of the metamaterial and only the nine unit cells directly above are folded. As the actuation magnet cells translate upward, as shown in positions 2, 3 and 4, unit cells no longer directly above the actuation magnet cells revert to the deployed mode near instantaneously. Meanwhile the unit cells directly above continue to actuate to the folded mode, demonstrating a rapid response to the dynamic movement of actuation magnet cells. Dynamic selective actuation is also achieved through rotation, as shown in **Figure 5C**. 25 actuation magnet cells arranged in a 5 × 5 grid rotate counterclockwise about the center point, with $\theta$ representing the rotation angle. The resulting experimental actuation from rotating actuation magnet cells is shown in **Figure 5D** and Video S5 (Supporting Information). At $\theta = 0^{\circ}$, the unit cells directly above the actuation magnet cells are folded, in a 5 × 5 pattern in the center of the metamaterial. Subsequently, the actuation magnet cells are rotated 90º and the unit cells begin to transition from the folded to deployed modes. Meanwhile, the unit cells in the upper right and lower left corners partially fold because of the rotating magnetic field. When $\theta = 180^{\circ}$, the unit cells directly above the actuation magnet cells are deployed but the rows above and below the deployed unit cells are slightly folded. Lastly, at $\theta = 270^{o}$, unit cells along a diagonal line starting from just inside the upper left corner and ending before the lower right corner. This inhomogeneous shape configuration is the opposite counterpart, in terms of deployed and folded unit cells, as the $\theta = 90^{o}$ configuration. Dynamic selective actuation represents another method to selectively actuate unit cells within a larger metamaterial for enhanced programmability and inhomogeneous shape transformation, with translation and rotation being demonstrated. Furthermore, it poses possibilities for study into the phase transition behavior between the deployed and folded modes.

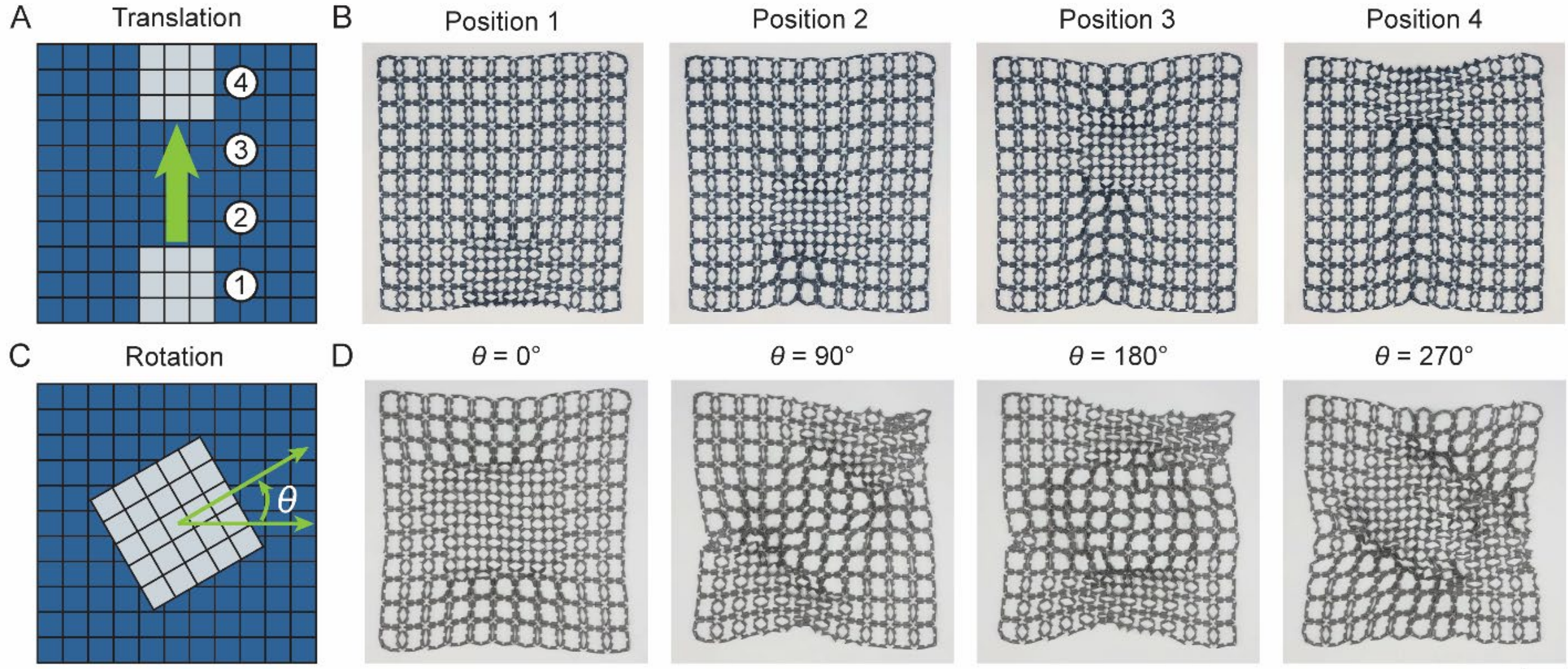

**Figure 5.** Dynamic selective actuation of magneto-mechanical metamaterials achieved by moving actuation magnet cells. (A) Schematic of a 3 × 3 grid of actuation magnet cells. The gray blocks represent the positions of the actuation magnet cells, and four positions of interest are marked as 1 through 4. (B) Experimental demonstration of the dynamic selective actuation of translating actuation magnet cells. (C) Schematic of the rotation path of a 5 × 5 grid of actuation magnet cells. The actuation magnet cells rotate counterclockwise about the center of the metamaterial and $\theta$ denotes the rotation angle of the actuation magnet cells. (D) Experimental demonstration of the dynamic selective actuation of rotating actuation magnet cells. Scale bar: 30 mm.

### 2.6. Magnetic selective actuation of flipped bilayer magneto-mechanical metamaterial

In a recent study, we demonstrated that a flipped bilayer magneto-mechanical metamaterial can preserve its overall area during actuation, i.e., folding of the units would not change the overall dimension of the metamaterial.[36] Here, selective actuation on the flipped bilayer system is presented. As illustrated in **Figure 6A,** the unit cell of the flipped bilayer metamaterial[36] consists of a 2 mm thick top layer and a 4 mm thick bottom layer. The magnetization directions between the top and bottom layers are flipped, allowing for magnetic attraction and therefore, good adhesion between the two layers. The flipped bilayer has three deformation modes (**Figure 6B**), which are the initial mode under no external magnetic field, the bottom folded mode under a downward magnetic field, and the top folded layer under an upward magnetic field. The three modes are represented by a navy block labeled as the "0" region, a gray block labeled as the "1" region, and an orange block labeled as the "-1" region, respectively. The multimodal magnetic programming is illustrated in **Figure 6C**. In the "0" region, no actuation magnet cell is plugged into the pegboard. To program a flipped bilayer unit cell into the bottom folded mode, a magnet cell is oriented such that the magnetization direction is pointed upward. Alternatively, to achieve the top folded mode, the magnetization direction of the magnet cells is pointed downward. Following this magnetic programming scheme, three selectively actuated patterns of the flipped bilayer metamaterial predicted by FEA simulation and experimental validation are demonstrated in **Figures 6D-F** (experimental actuation is also provided in Video S6). Enlarged views of the unit cells in the red boxes are presented for better comparison. For the selective bottom folding pattern shown in **Figure 6D**, the bottom unit cells are selectively actuated to form an "X" pattern. As is the feature of the flipped bilayer, the overall shape of the metamaterial remains nearly unchanged. For the selective top folding pattern (**Figure 6E**), the centermost 16 unit cells arranged in a $4 \times 4$ grid are in the top folded mode. For the combined folding pattern (**Figure 6F**), the top right and bottom left corners are top folded and the top left and bottom right corners are bottom folded. As can be seen, the FEA predictions agree with the experimental results very well for all the three cases. This demonstrates that the selective actuation strategy for inhomogeneous deformations also applies to flipped bilayer magneto-mechanical metamaterials. Therefore, the multifunctionalities in elastic wave manipulation achieved in the single-layer magneto-mechanical metamaterial (demonstrated in sections 2.2 to 2.4) such as elastic wave transmittance tuning, elastic waveguide, and vibration isolation, can also be extended to the flipped bilayer metamaterials.

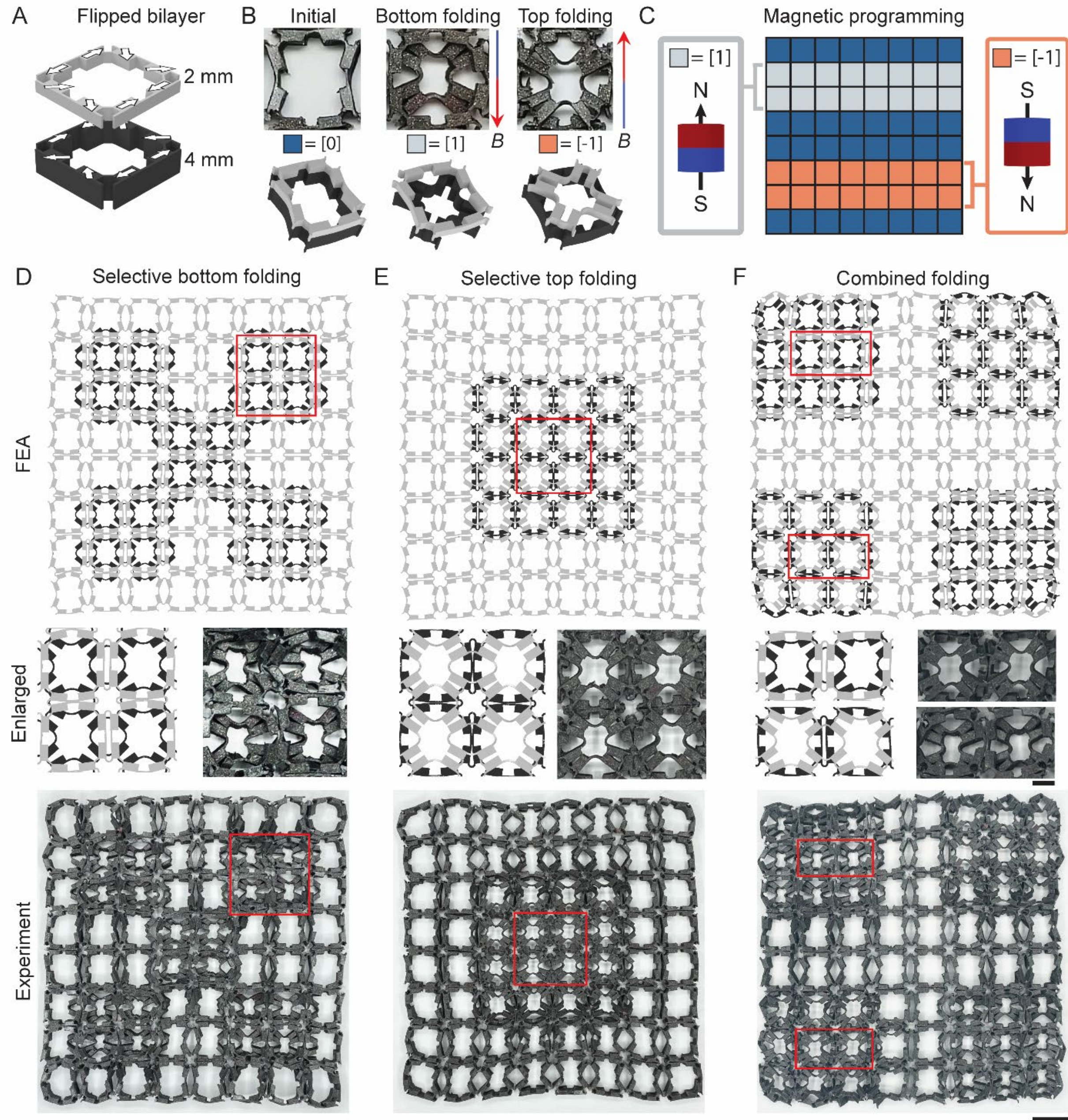

**Figure 6**. Multimodal selective actuation of the flipped bilayer magneto-mechanical metamaterial. (A) Unit cell design of the flipped bilayer metamaterial. The 2 mm thick top layer stacks on a 4 mm thick bottom layer, and the magnetization direction between the two layers are flipped. (B) The three deformation modes of the flipped bilayer and their corresponding color, number code in the magnetic programming and illustrated configuration. Scale bar: 5 mm. (C) Actuation magnet cell orientations for a magnetic programming example. (D-F) Three selectively actuated patterns of an 8 × 8 flipped bilayer metamaterial predicted by FEA with experimental validation. (D) Selective bottom folding pattern where the bottom unit cells are folded in an "X" pattern. (E) Selective top folding pattern where the centermost 4 × 4 top unit cells are folded. (F) Combined folding pattern in which both top folded, top right and bottom left corner, and bottom folded, top

left and bottom right corner, unit cells are included in selective combined folding. Scale bars in the enlarged view is 5 mm and in the bottom row is 20 mm.

## 3. Conclusion

In summary, we have presented a selective actuation strategy for inhomogeneous deformation of magneto-mechanical metamaterials, which is enabled through an actuation magnet cell wherein permanent magnets actuate individual unit cells. By programming the arrangement of actuation magnet cells, the metamaterial can achieve diverse inhomogeneous shape configurations through selective actuation of its unit cells between the deployed and folded shape modes. The proposed selective actuation strategy significantly increases programmability of magneto-mechanical metamaterials for elastic wave propagation functions, such as transmittance tuning, elastic waveguides, vibration isolation, and direction-dependent elastic wave propagation control, all integrated into a single multifunctional metamaterial. Compared to traditional heavy and bulky electromagnetic coils used for magneto-mechanical metamaterial actuation, the carefully designed actuation magnetic cells are lightweight, compact, and highly portable. They are also capable of actuating the metamaterial across distances, even when separated by bulk material or when the metamaterial is encapsulated inside a shell-like structure. Consequently, it becomes straightforward to achieve dynamic selective actuation by simply translating or rotating the actuation magnet cells. The programming and reprogramming of selective actuation patterns can be further enhanced by the automation of the positioning of the actuation magnet cells. Although we only demonstrate tuning elastic wave transmittance and propagation in this work, the selective actuation strategy can be used to program other properties and behaviors of magneto-mechanical metamaterials, such as stiffness, Poisson's ratio, and stress-strain response. Furthermore, the unit cells of the magneto-mechanical metamaterial in this work are wholly fabricated with a soft magnetic composite material. However, the design is not limited to soft materials and can be replicated with an appropriate combination of rigid blocks and soft hinges. Such a unit cell design reduces damping effects in practical applications and enables experimental studies to explore the full potential of elastic wave manipulation in more rigid selectively actuated metamaterials. Other materials may also be utilized to target higher or lower frequency ranges. In addition, while this work demonstrates tuning elastic wave transmittance and propagation in the final deformed state of the selective actuation pattern, selective actuation can be further investigated by studying the dynamic changes in both transmittance and propagation as the metamaterial undergoes selective actuation. Understanding how the wave manipulation property evolves during the transition between selective actuation patterns can create new opportunities for real-time control of wave behavior. We anticipate that the selective actuation strategy can serve to widen the application space of magneto-mechanical metamaterials and inspire the design of new multifunctional active metamaterials whose functionality and application can be readily adjusted by transforming the structure into new configurations.

## 4. Experimental Section

*Metamaterial fabrication*: Dragon Skin 20 (Smooth-On Inc., Macungie, PA, USA) and NdFeB particles (average particle size of 100 μm, Magnequench, Singapore) were mixed at a volume ratio of 3:1 and then degassed in a vacuum chamber to remove entrapped air. The mixture was then injected into Ultimaker S5 (Ultimaker, Netherlands) 3D-printed polyvinyl alcohol (PVA) molds. The filled molds were sandwiched between two glass slides and set to cure at 80 ºC for 30 minutes. After the molds were removed, the unit cell was magnetized under a 1.5 T magnetic impulse field. Magnetized unit cells were then connected using Sil-poxy adhesive (Smooth-On, Inc., Macungie, PA, USA) until an 11 × 11 metamaterial array was fabricated. This fabrication process was adapted from a previous study.[36] More details are provided in the Supporting Information.

*Structural actuation setup*: The permanent magnet used in this study was an axially magnetized permanent N52 NdFeB ring magnet. Four magnets were arranged in a stack such that the magnetization direction of all the magnets were aligned. The magnetic stack was then embedded inside a 3D-printed polylactic acid (PLA) fixture. The fixture included a square peg on the bottom to allow it to be plugged into a corresponding 3D-printed PLA pegboard. Additional PLA fixtures were 3D-printed to firmly secure the actuation magnet cells in place. Further information on the magnets, actuation magnet cell, and pegboard can be found in Figure S1 (Supporting Information).

*Structural deformation simulation*: The structural deformation of the metamaterial was predicted using the commercial FEA software ABAQUS 2021 (Dassault Systèmes SE, Vélizy-Villacoublay, France) and a custom user element subroutine for hard-magnetic soft materials.[50] Arrays of deployed unit cells were assigned a shear modulus, bulk modulus, and magnetization. An external applied magnetic field magnitude was applied to the metamaterial arrays and surface-to-surface contact with a friction coefficient of 0.3 was used to prevent surface penetration. Figure S5 (Supporting Information) includes more information on the structural simulation.

*Transmittance curve calculation*: The transmittance curves were obtained using ABAQUS 2021. Selective actuation configurations were exported from the structural simulations and a steady-state dynamic analysis was performed on the configurations for the transmittance-frequency curves. A specific side's nodes were multipoint coupled (MPC) to an excitation node and the opposite side's nodes were likewise MPC constrained to a listener node. The transmittance curves were then calculated by analyzing the displacement amplitude of the listener node. More information on the transmittance curve calculation can be found in the Supporting Information.

*Elastic wave propagation simulation*: Selectively actuated configurations were exported from the structural simulation and then a dynamic, explicit step was used to simulate the elastic wave propagation in ABAQUS 2021. The elastic wave frequency and number of excitations were defined by the circular frequency and the total step time. A displacement magnitude contour was used to visualize the elastic wave. More information on the elastic wave propagation simulation can be found in the Supporting Information.

**Supporting Information**

Supporting Information is available from the Wiley Online Library or from the author.

**Acknowledgements**

The authors acknowledge the support from NSF Career Award CMMI-2145601 and NSF Award CMMI-2142789.

**Conflict of interest**

The authors declare no conflict of interest.

**Data Availability Statement**

The data that support the findings of this study are available from the corresponding author upon reasonable request.

**Author Contributions**

R.R.Z. designed the research; J.S., S.W., and S.H. performed the research; J.S. and S.H. carried out the experiments; J.S. and S.W. conducted the FEA simulations; All authors wrote the paper.